\documentclass[twocolumn,showpacs]{revtex4}
\usepackage{amssymb,epsfig,amscd}
\usepackage{graphicx}% Include figure files
\usepackage{dcolumn}% Align table columns on decimal point
\usepackage{bm}% bold math

\usepackage{epsfig}

\def\be{\begin{equation}} 
\def\ee{\end{equation}}
\def\bq{\begin{eqnarray}} 
\def\eq{\end{eqnarray}}
\def\de{\delta}
\def\al{\alpha}
\def\bt{\beta}

\begin{document}

\title{Bulk Viscosity in Hybrid Stars}
\author{A. Drago$^a$, A. Lavagno$^b$, G. Pagliara$^a$}
\address{$^a$Dipartimento di Fisica, Universit{\`a} di Ferrara and
INFN, Sezione di Ferrara, 44100 Ferrara, Italy\\
\noindent
%\address{
$^b$Dipartimento di Fisica, Politecnico di Torino and INFN,
Sezione di Torino, 10129 Torino, Italy}

%\date{\today } 

\begin{abstract} 

We compute the bulk viscosity of a mixed quark-hadron phase.  In the
first scenario to be discussed, the mixed phase occurs at large
densities and we assume that it is composed of a mixing of hyperonic matter
and quarks in the Color Flavor Locked phase.  In a second scenario,
the mixed phase occurs at lower densities and it is composed of a
mixing of nucleons and unpaired quark matter. 
We have also investigated the effect of a non-vanishing
surface tension at the interface between hadronic and quark matter.
In both scenarios, the bulk viscosity is large when the surface tension is absent,
while the value of the viscosity reduces in the second scenario when
a finite value for the surface tension is taken into account. In all cases,
the r-mode instabilities of
the corresponding hybrid star are suppressed.
\end{abstract}


\noindent

\maketitle

\section{Introduction}
The discovery by Andersson, Friedman and Morsink of r-mode
instabilities in neutron stars put rather severe limits on the
highest rotation frequency of pulsars \cite{Andersson:1998xt,Friedman:1998uh}.
These constraints can be incompatible with the existence of
millisecond pulsars, if the instability is not suppressed by a
sufficiently large viscosity.  Actually, for a star composed only of
neutrons and protons and for temperatures larger than roughly
10$^{10}$ K, the bulk viscosity due to the modified Urca process is
large enough to damp the instability \cite{Sawyer:1989}.  On the other
hand, when the star cools down to lower temperatures, the instability
is not suppressed and the star is forced to loose angular momentum via
emission of gravitational waves \cite{Lindblom:1999yk}.  More recently, it has been noticed
that the existence of viscous boundary layers at the interface between
the fluid core and the crust can stabilize the star \footnote{Also the
presence of a magnetic field inside the compact star can damp the
r-modes on a time-scale of order hours or
days~\cite{Rezzolla:1999he}.} if the rotation period is longer than
$\sim$ 1.5 ms \cite{Bildsten:1999zn,Andersson:2000pt}.  At very low
temperatures, below 10$^8$ K, shear viscosity becomes large and it
allows older stars to increase their angular velocity by mass
accretion.

Actually, compact stars can be constituted by a larger variety of particles than
just neutrons and protons. One possibility is that hyperons form at the
center of the star. It has been shown that, due to non-leptonic weak reactions,
bulk viscosity can be rather large for an hyperonic 
star \cite{Jones:2001ya,Lindblom:2001hd},
which therefore can emit gravitational waves only if its temperature
is $\sim 10^{10}$K and its frequency is larger than 10--30\% of its
Keplerian frequency. 

The formation of quark matter inside a compact star has been discussed
extensively in the literature.  Bulk viscosity of non-interacting
strange quark matter is very large
\cite{Wang:1984tg,Sawyer:1989uy,Madsen:1992sx}.  On the other hand,
recent studies taking into account quark-quark interaction revealed
the possible existence of color superconducting phases in which quarks
form Cooper pairs with gaps as large as 100 MeV
\cite{Rajagopal:2000wf}. In that case, bulk viscosity is strongly
suppressed by the large energy gaps, r-mode instabilities are not
damped and pure color-superconducting quark stars seem therefore to be
ruled out by the pulsar data \cite{Madsen:1999ci}.

In the present paper we are interested in studying the viscosity and
stability, respect to r-modes, of an Hybrid Star (HyS), for which no
quantitative analysis has been performed so far.  If the star is made 
of hyperonic matter and of non-interacting quarks, it is rather
obvious that the viscosity of the HyS should be large, due to the
large viscosity of its constituents. The real question concerns the
case in which color-superconducting quark matter is considered, since
its viscosity is negligible. There are in principle (at least) two
sources of viscosity for HySs.  One originates at the
interface between the crust and the fluid interior
\cite{Madsen:1999ci} and it is similar to the one discussed above in
the case of purely hadronic stars
\cite{Bildsten:1999zn,Andersson:2000pt}. In our work we will not
discuss that possibility and we will instead concentrate on a
quantitative evaluation of the bulk viscosity in a mixed phase (MP) 
composed either of hyperons and color-superconducting quark matter 
or of nucleons and non-interacting quarks.  
We have also investigated the effect of the existence of finite-size
structures in the MP. These geometrical structures can exist if
the surface tension at the interface between hadronic and quark matter is
not vanishing. 
We will show that the
bulk viscosity of the MP is in general rather large, particularly so when the surface
tension is negligible. Therefore HySs are possible 
candidates for young millisecond pulsars.

\section{Equation of State}
We construct the Equation of State (EOS) of matter at high density
modeling the Hadronic phase by a relativistic non-linear Walecka type
model \cite{Liu:2001iz} with the inclusion of Hyperons
\cite{Glend:01kn}. Concerning quark matter, we use an MIT-bag like EOS
in which quarks can pair to form a Color Flavor Locked (CFL) phase
\cite{Alford:1998mk,Alford:2001zr,Alford:2002rj}. CFL is considered to
be the energetically favored type of pairing pattern, at large
densities, in the case of $\beta$-stable, electric-charge neutral
quark matter.  At lower densities, a two-flavor Color Superconducting
(2SC) phase can form, with a smaller energy gap \cite{Neumann:2002jm}.

Two scenarios are possible, depending on the value of the critical
density separating hadronic matter from MP.  In the first case the
critical density is large, therefore hyperons start being produced
at a density smaller than the critical one. We will call this phase
``hyperon-quark'' MP \cite{Banik:2002kc}. This is the first case
discussed below and it corresponds to the upper panel of Fig.~\ref{densita}. In
the second scenario, the MP starts at a density lower than the
hyperonic threshold and the hyperon density is much smaller than in
the previous case. We will explore in particular the situation in
which hyperons are completely eaten-up by quarks and they do not
appear in the ``nucleon-quark'' MP (see lower panel of Fig.~\ref{densita}). In
this second scenario we will assume that in the MP the density is so
low that the CFL pairing cannot form and, for simplicity, we will
consider unpaired quarks in the MP, since anyway the bulk viscosity of
the 2SC quark phase is similar to the viscosity of unpaired quark
matter \cite{Madsen:1999ci}. In both scenarios, we describe the
composition of the MP assuming a first order transition and imposing
Gibbs conditions \footnote{The structure of the MP obtained imposing Gibbs condition 
is clearly 
different from a two-fluid model like the one
adopted e.g. in Ref. \cite{Andersson:2002jd}, in which the two phases are essentially 
independent}. 
Beta stability and charge neutrality are satisfied
in every density region.

\section{Bulk Viscosity}
Bulk viscosity is the dissipative process in which a perturbation of
the pressure of a fluid element is converted to heat.  A small
variation of the pressure of the system can be treated as a
perturbation on the densities of the different species of particles
bringing the system out of $\beta$-equilibrium. The reactions between
the different particles drive the system back to an equilibrium
configuration, with a delay which depends on the characteristic time
scale of the interactions.  Following the formalism of
Ref.\cite{Lindblom:2001hd}, we classify all the reactions either as
fast or slow.  Slow processes produce bulk viscosity, because their
time scale is comparable with the period of the perturbation.  Fast
processes, instead, put constraints on the variation of the densities
of the different particles. In the following we generalize the
formalism of Ref. \cite{Lindblom:2001hd} in order to compute the bulk
viscosity of MP.  The equations needed in the computation are rather
different in the two scenarios discussed above and we will have to
deal with all the different cases separately. Moreover, if the effect
of a non-vanishing surface tension is taken into account,
the formalism needed to compute the 
bulk viscosity has to be further modified.  
It is well known that a finite value for the surface tension allows
the formation of finite size structures whose geometrical shape is 
determined by the interplay of the various contribution to the energy,
including the Coulomb term \cite{Heiselberg:1992dx,Voskresensky:2002hu}. 
A precise estimate of surface tension $\sigma$ is unfortunately still
lacking, and values ranging from a few MeV/fm$^2$ to a few tens
MeV/fm$^2$ have been discussed in the literature.  We can roughly
divide this range of values in three windows.  Values larger than $\sim$
30 MeV/fm$^2$ would not allow the formation of the mixed phase since
it would not be energetically favored\cite{Alford:2001zr,Voskresensky:2002hu}. 
Our analysis is therefore
restricted to smaller values.  In the following we will discuss both
the case in which the surface tension is so small that it can be neglected
and the case in which it is non-vanishing.

\subsection{Negligible surface tension}
Let us first discuss the case in which the surface tension is so small
that the perturbations of the star, associated e.g. to r-modes, can
break the finite-size structures present in the MP.  We have estimated
that this case corresponds to values of the surface tension $\sigma
\lesssim $1 MeV/fm$^2$, since to r-modes excitations is associated an
energy per baryon of a few MeV.

\subsubsection{First scenario}
\begin{figure}[]
\begin{center}
\includegraphics[scale=0.55]{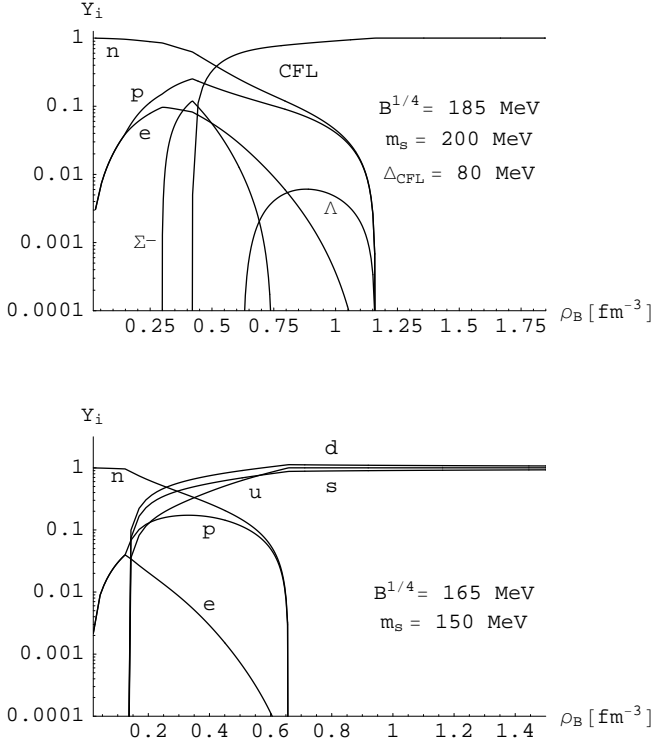}
\end{center}
\parbox{7.5cm}{
\caption{\label{densita}
Particle abundances, as function of the total baryon density. 
The upper panel corresponds to the case of
hyperon-quark MP and the lower panel to nucleon-quark MP.}
}
\end{figure}

\begin{figure}[]
\begin{center}
\includegraphics[scale=0.55]{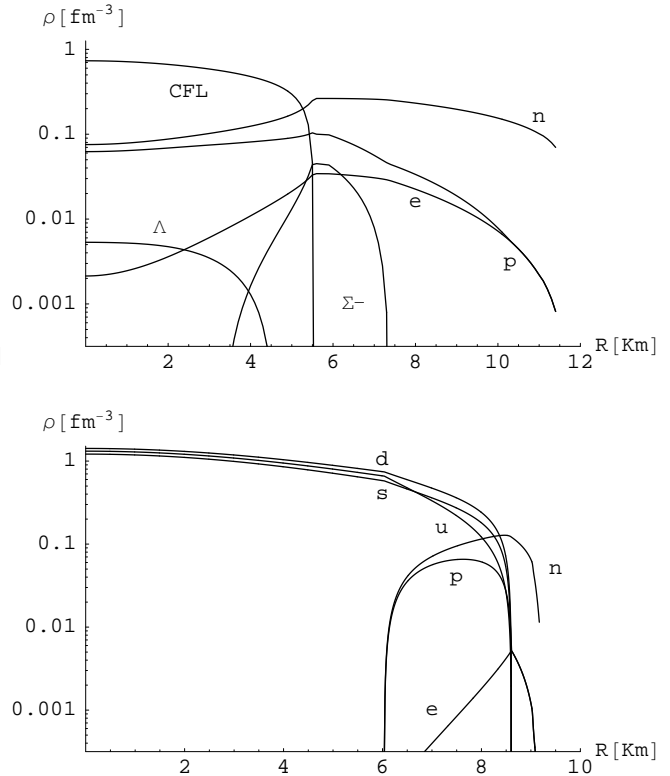}
\end{center}
\parbox{7.5cm}{
\caption{\label{profili}
Particles densities profiles inside hybrid stars 
of mass $M=1.46 M_\odot$ in the two discussed scenarios.
Parameters as in Fig.~1.}
}
\end{figure}

We start our analysis from the first scenario discussed in
Sec. II, namely from the case in which
hyperons are present in the MP.  We will discuss in particular a
parameter set in which only $\Lambda$ and $\Sigma^-$ particles are
produced in the MP. The only slow processes generating viscosity are
non-leptonic reactions between hadrons \footnote{We assume, as in
Ref.\cite{Lindblom:2001hd}, that all leptonic reaction rates are much
smaller than those associated with non-leptonic reactions, so we do
not include them in the calculation of bulk viscosity of the MP. The
viscosity due to semi-leptonic reactions (modified Urca processes)
becomes relevant at very high temperatures and we will include its
effect when computing the stability of the star.}, since the large gap
prevents weak reactions between quarks. As in
Ref.\cite{Lindblom:2001hd}, we consider the reactions
\bq 
n+n &\stackrel{H_W}{\longleftrightarrow}& p+\Sigma^-\, ,\label{nnps}\\ 
n+p &\stackrel{H_W}{\longleftrightarrow}& p+\Lambda\, .\label{nppl}
\eq 
Following \cite{Lindblom:2001hd}, we have not taken into account
the reaction $n+n \stackrel{H_W}{\longleftrightarrow}n+\Lambda$,
since the corresponding reaction rate cannot be easily estimated.
In Ref.\cite{Jones:2001ya} it has been argued that this rate can be
one order of magnitude larger that the one associated with reaction (\ref{nppl}). 
In the first scenario, our results have therefore to be considered as 
upper limits for the viscosity, as it will be clarified in the following.
Notice anyway that, concerning the damping of r-modes instabilities,
to be discussed in the following, the most important region of the star
corresponds to low--moderate densities, while the $\Lambda$ is produced
at larger densities as shown in Figs.~(\ref{densita}) and (\ref{profili}).

Concerning fast processes, they come from the following reactions
mediated by the strong interaction
\bq 
n+\Lambda &\stackrel{H_S}{\longleftrightarrow}& p+\Sigma^-\, ,\\ 
\Lambda+\Lambda &\stackrel{H_S}{\longleftrightarrow}& 2(uds)_{CFL}\,\label{melting} . 
\eq
The last process describes the only possible ``melting'' of hadrons
into CFL phase, since the pairing forces the number of up, down and
strange quarks to be equal.  
This process is fast because of the vanishing value of $\sigma$.
As we will see later the ``melting'' process is instead forbidden
if the value of $\sigma$ is not negligible.  
Concerning mechanical equilibrium, elastic scattering due to strong
interactions, as well as melting processes like the one
described in Eq.(\ref{melting}), are responsible for (rapid) momentum transfer between the
two phases. 
The mechanical equilibrium between the two phases 
is reached in a time scale $t_s$ determined by  
strong interaction, $t_s \sim 10^{-23}$ s. The full system is, instead, 
out of mechanical equilibrium
on a time scale of the order of the period of the perturbation $t_p \sim 10^{-3}$ s. 
Therefore, during a fluctuation the two components of the fluid remain
in mutual mechanical equilibrium. 
The variations of the pressure in the two
phases have therefore to satisfy the constraint:
\be 
\de P_h=\de P_{CFL} \, .
\ee

%%%%%%%%%%%%%%%%%%%%%%%%%%%%%%%%%%%%%%%%%%%%%%%%%%%%%%%%%%%%%%%%%%%%%%%%%%%%%%%%%%%%%%%%%%%%%

%%%%%%%%%%%%%%%%%%%%%%%%%%%%%%%%%%%%%%%%%%%%%%%%%%%%%%%%%%%%%%%%%%%%%%%%%%%%%%%%%%%%%%%%%%%%%%

The variations $\de\rho_i$ of the densities of the various particles
and the variation $\de\chi$ of the quark fraction are constrained by
the following linearized equations
\bq
0 &=& (1-\chi)(\de\rho_n+\de\rho_p+\de\rho_\Lambda+\de\rho_\Sigma)\nonumber\\
&+&\chi(\de\rho_q)+\de\chi(\rho_q-\rho_n-\rho_p-\rho_\Lambda-\rho_\Sigma)\, ,\\
0 &=& (1-\chi)(\de\rho_p-\de\rho_\Sigma)-\de\chi(\rho_p-\rho_\Sigma)\, ,\\
0 &=& \sum_{\{H\}}p_H\de\rho_H- p_q\de\rho_q\, ,\\
0 &=& \beta_n\de\rho_n+\beta_p\de\rho_p+\beta_\Lambda\de\rho_\Lambda+\beta_\Sigma\de\rho_\Sigma\, ,\\
0 &=& \al_{\Lambda n}\de\rho_n+\al_{\Lambda p}\de\rho_p+\al_{\Lambda \Lambda}\de\rho_\Lambda\nonumber\\
&+&\al_{\Lambda \Sigma}\de\rho_\Sigma-3\al_{q q}\de\rho_q\, ,
\eq
where
\bq
\al_{i j} &=& \left(\partial\mu_i/\partial\rho_j\right)_{\rho_k,k\neq j}\, ,\\
\bt_i &=& \al_{n i}+\al_{\Lambda i}-\al_{p i}- \al_{\Sigma i}\, ,\\
p_i &=& \partial P/\partial \rho_i\, .
\eq

\noindent
Eqs.~(6),(7) impose baryon number conservation and electric
charge neutrality.  Eq.~(8) (where the sum runs on all hadrons)
imposes the mechanical equilibrium defined by Eq.~(5).  Finally,
Eqs.~(9),(10) describe the equilibrium respect to the two strong
processes of Eqs.~(3),(4).  
Notice that $\de\chi$ does not appear in Eqs.~(8)--(10),
because neither the pressure nor the chemical potentials
explicitely depends on the quark volume fraction $\chi$.
Solving the system allows to express all
the $\de\rho_i$ and $\de\chi$ as function of $\de\rho_n$ \footnote{\label{nota}If
the $\Sigma$ density vanishes, the corresponding density fluctuation
is identically zero and the constraint given by Eq.~(3) has not to be
imposed. If the $\Lambda$ density vanishes, then $\de\rho_\Lambda=0$,
Eq.~(3) has not to be imposed, the melting process described by
Eq.~(4) does not exist and $\de\chi=0$.}.

%%%%%%%%%%%%%%%%%%%%%%%%%%%%%%%%%%%%%%%%%%%%%%%%%%%%%%%%%%%%%%%%%%%%%

\begin{figure}[]
\begin{center}
\includegraphics[scale=0.57]{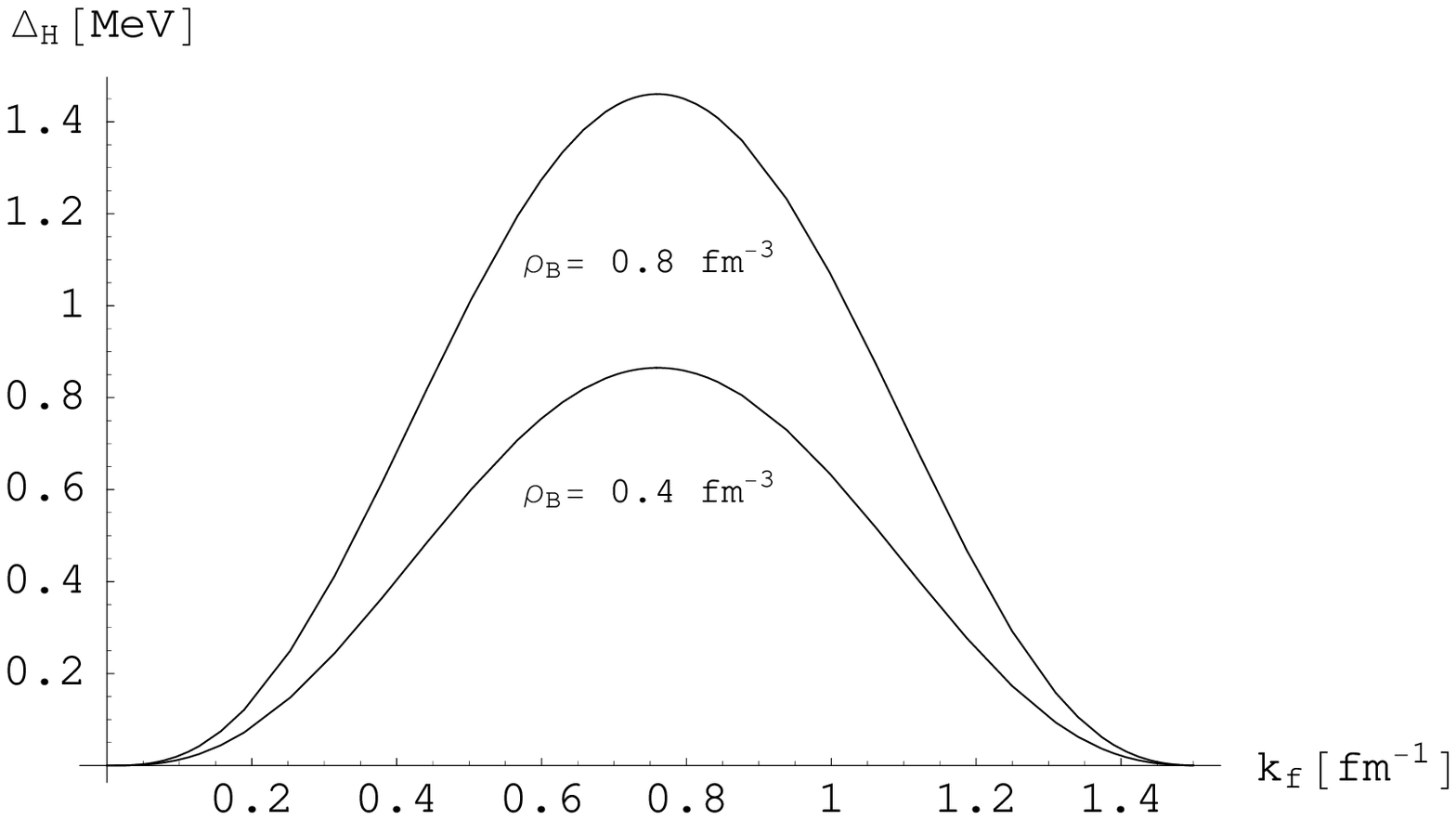}
\end{center}
\parbox{7.5cm}{
\caption{\label{gappi} Hyperon superfluid gap at zero temperature 
and for two values of the  baryon density. (see Ref.~\cite{Lindblom:2001hd})  }  }
\end{figure}

%%%%%%%%%%%%%%%%%%%%%%%%%%%%%%%%%%%%%%%%%%%%%%%%%%%%%%%%%%%%%%%%%%%%

%%%%%%%%%%%%%%%%%%%%%%%%%%%%%%%%%%%%%%%%%%%%%%%%%%%%%%%%%%%%%%%%%%%%%

\begin{figure}[]
\begin{center}
\includegraphics[scale=0.57]{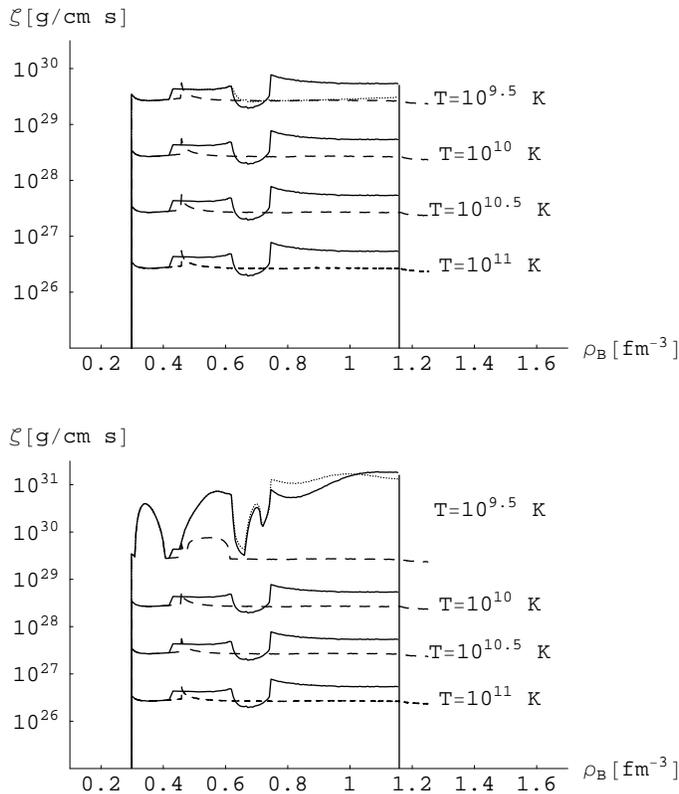}
\end{center}
\parbox{7.5cm}{
\caption{\label{vis1}
Bulk viscosity in the first scenario as a function of baryon density for
various temperatures. Solid lines refer to MP viscosity, while dashed
lines correspond to pure hyperonic matter (vanishing hyperon gap in the upper panel
and finite hyperon gap in the lower panel).
The dotted lines corresponds to the case in which a large
hadron-quark
surface tension (10 MeV/fm$^2$  $< \sigma <$ 30 MeV/fm$^2$) is taken into account and they are computed for
T$=10^{9.5}$K (see text).}  }
\end{figure}

%%%%%%%%%%%%%%%%%%%%%%%%%%%%%%%%%%%%%%%%%%%%%%%%%%%%%%%%%%%%%%%%%%%%
The relaxation time $\tau$ associated to the weak processes reads
\be
\frac{1}{\tau}=\left(\frac{\Gamma_\Lambda}{\de\mu}+2\frac{\Gamma_\Sigma}
{\de\mu}\right)\frac{\de\mu}{\de\rho_n}\, .\label{relax}
\ee
Here $\Gamma_\Lambda$ and $\Gamma_\Sigma$ are the rates of the weak
interactions \footnote{Notice that the Coulomb interaction does not play any direct role
when computing the viscosity in the present scheme, since it cannot modify the 
value of the chemical unbalance $\delta \mu$.}.
They have been calculated in 
Eqs.~(4.21), (4.28), (4.29) of
Ref.\cite{Lindblom:2001hd},
where it has been shown that the main dependence of the rate on the temperature
is a quadratic one. Moreover, if hyperon superfluid gaps as the ones displayed in Fig.~\ref{gappi} 
do develop, they exponentially suppress the rates as it will be discussed later.

The chemical potential
unbalance $\de\mu$ is given by
\be
\de\mu \equiv \de\mu_n-\de\mu_\Lambda=2\de\mu_n-\de\mu_p-\de\mu_\Sigma\, ,
\ee
where the variations of the 
chemical potentials can be expressed in terms of $\de\rho_i$ as
$
\de\mu_i =\sum_j \al_{i j}\de\rho_j\, .
$
A unique $\de\mu$ appears in Eq.~(15), since the variations
$\de\mu_\Lambda$ and $\de\mu_\Sigma$ are constrained by the fast
process given in Eq.~(3).  The real part of bulk viscosity, which is the
relevant quantity for damping r-modes instabilities, finally reads
\be
\zeta = \frac{P (\gamma_\infty-\gamma_0)\tau}{1+(\omega \tau)^2} \, ,
\ee
where $\gamma_\infty$ and $\gamma_0$ are the ``infinite'' frequency
adiabatic index and the ``zero'' frequency adiabatic index and
$\omega$ is the angular velocity of the perturbation.

It is interesting to remark that there are two asymptotic behaviors of the
viscosity as a function of the frequency $\omega$. In the high
frequency case ($\omega\tau\gg 1$), the viscosity scales as $1/\tau$,
while in the low frequency limit the viscosity is proportional to
$\tau$. In the parameter and temperature ranges explored in our paper
it results that we always remain in the low frequency limit, as it can
be seen from Fig.~\ref{vis1} noticing that the viscosity decreases with the
temperature. In this regime the addition of another weak decay
(e.g. $n+n \stackrel{H_W}{\longleftrightarrow}n+\Lambda$) decreases 
the relaxation time and as consequence the viscosity too.
In this sense our results for the viscosity in the first scenario
must be considered as upper limits.

We can observe from Fig.~\ref{vis1} that, in the
``hyperon-quark'' MP scenario discussed so far, MP bulk viscosity is comparable to bulk viscosity of
purely hyperonic matter even though the density of hyperons is very
small in the MP (see upper panel of Fig.~\ref{densita})
\footnote{Our results in both scenarios are consistent with the
general outcome of Ref.\cite{Haensel:2001mw} stating that in the
high-frequency limit the bulk viscosity is just the sum of partial
bulk viscosities in different slow reaction channels.}. 
Notice that in the small window of baryonic densities in which
both $\Sigma$ and $\Lambda$ hyperons are present, the value of bulk viscosity 
of the MP and of the pure hadronic phase are essentially identical. On the other
hand, in the  baryonic density  windows in which only one species of hyperons
are present in the MP, the bulk viscosity of MP is larger than the bulk viscosity
of the pure hadronic phase because only one decay channel 
is open ($\Sigma$ decay of Eq.~(\ref{nnps}) or $\Lambda$ decay of Eq.~(\ref{nppl})) 
and therefore the relaxation time is larger.
It is interesting to compute the bulk viscosity in this scenario also including
the effect of hyperon superfluidity, as done in Ref.~\cite{Lindblom:2001hd}.
If the energy gaps $\Delta_H$ associated with the hyperons are non vanishing, 
the decay rates are suppressed by the factor $e^{-\Delta_H/T}$ where the gaps 
have a typical shape shown in Fig.~\ref{gappi}. For low temperatures 
the viscosity displayes the characteristic features shown in the lower panel of 
Fig.~\ref{vis1}, while these features disappear at larger temperatures or 
for vanishing $\Delta_H$.  
Concerning the appreciable difference  
between the viscosity of MP and of pure hadronic matter for $T=10^{9.5}$K, 
this is due to the Fermi momentum 
dependence of $\Delta_H$, which implies that the gaps are suppressed for large hyperon densities.
In the MP, the density of hyperons is lower than in pure hadronic matter and 
therefore the effect of the gaps shows out more dramatically.
 
Let us briefly discuss another possible source of bulk viscosity
related to the formation of a boson condensate.
Since all quarks in the CFL phase are gapped, the low energy excitations
are the Nambu-Goldstone bosons associated with the spontaneous 
symmetry breaking of the global symmetries \cite{Alford:1998mk}.
In particular, a condensate of $\pi^{-}$ or $K^{-}$ can appear, in the MP,
as proposed in Ref.~\cite{Alford:2002rj}.  
The role played by these boson condensates in the cooling
of a CS has been studied in Ref.~\cite{Jaikumar:2002vg}
and the condensates can play a role also in the calculation of the bulk viscosity of the MP. 
The weak processes involving these bosons are:
\bq
\pi^{-}&\stackrel{H_W}{\longleftrightarrow}& e^{-}+ \bar\nu_e\, ,\label{pioni}\\
K^{-}&\stackrel{H_W}{\longleftrightarrow}& e^{-}+ \bar\nu_e\, .\label{kaoni}
\eq
In the absence of these bosons, the variation of the density of electrons
$\de\rho_e$ can be neglected at low temperature ($T < 10^{10}$K)
because, as already remarked, all leptonic reaction rates are much
smaller than those associated with non-leptonic reactions.
If, on the other hand, the decay channels of Eqs.~(\ref{pioni}),(\ref{kaoni}) are open, 
then the corresponding decay rates are of the same order of the rates of the non-leptonic 
processes of Eqs.~(\ref{nnps}),(\ref{nppl}) and,
therefore, a new contribution to the viscosity appears. While a detailed calculation 
is clearly needed, the main result of our work, namely the existence of a large viscosity 
for $T \lesssim 10^{10}$ K, would even be strengthened.  

%%%%%%%%%%%%%%%%%%%%%%%%%%%%%%%%%%%%%%%%%%%%%%%%%%%%%%%%%%%%%%%%%%%%%

\begin{figure}[]
\begin{center}
\includegraphics[scale=0.57]{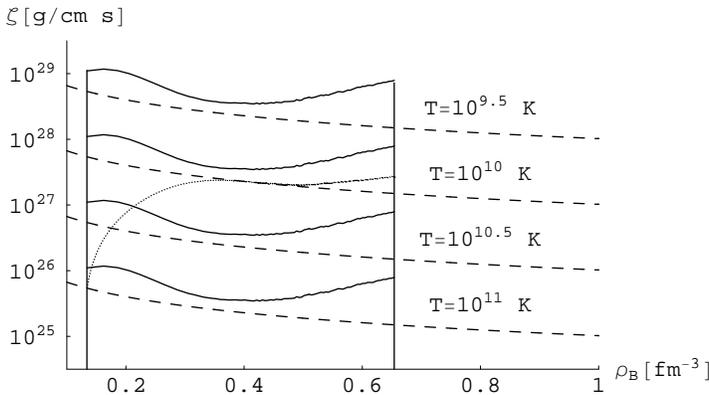}
\end{center}
\parbox{7.5cm}{
\caption{\label{vis2}
Bulk viscosity as a function of baryon density, for
various temperatures in the second scenario. Solid lines refer to MP viscosity, while dashed
lines correspond to pure quark matter.
The dotted line corresponds to the case in which a large
nucleon-quark
surface tension (10 MeV/fm$^2$  $< \sigma <$ 30 MeV/fm$^2$) is taken into account and it is computed for
T$=10^{9.5}$K (see text).}  }
\end{figure}

%%%%%%%%%%%%%%%%%%%%%%%%%%%%%%%%%%%%%%%%%%%%%%%%%%%%%%%%%%%%%%%%%%%%

\subsubsection{Second scenario}

We now discuss the second scenario in which a ``nucleon-quark'' MP
forms, hyperons are absent and CFL quark pairing cannot take place in
the MP \footnote{Notice that if CFL gaps can form at low density then
the bulk viscosity of MP vanishes.}.  We will assume that in the pure
quark matter phase CFL gaps can form so that this phase will not
contribute to the viscosity. The only source of viscosity, neglecting
as before semi-leptonic reactions, is therefore
\be
d+u \stackrel{H_W}{\longleftrightarrow} u+s\, .
\ee

%%%%%%%%%%%%%%%%%%%%%%%%%%%%%%%%%%%%%%%%%%%%%%%%%%%%%%%%%%%%%%%%%%%%%

%%%%%%%%%%%%%%%%%%%%%%%%%%%%%%%%%%%%%%%%%%%%%%%%%%%%%%%%%%%%%%%%%%%%

Concerning fast processes, they are given by the ``melting'' of
nucleons into unpaired quarks. The chemical equilibrium respect to
these reactions must therefore be satisfied during the perturbation.
The linearized equations governing the density fluctuations, analogous
to Eqs.~(6)-(10) of the first scenario, read
\bq
0 &=& (1-\chi)(\de\rho_n+\de\rho_p)+\chi(\de\rho_u+\de\rho_d+\de\rho_s)/3\nonumber\\
  &+& \de\chi\left((\rho_u+\rho_d+\rho_s)/3-\rho_n-\rho_p\right)\, ,\\
0 &=& (1-\chi)\de\rho_p+\chi(2\de\rho_u-\de\rho_d-\de\rho_s)/3\nonumber\\
  &+& \de\chi\left((2\rho_u-\rho_d-\rho_s)/3-\rho_p\right)\, ,\\
0 &=& \sum_{\{H\}}p_H\de\rho_H-\sum_{\{Q\}}p_Q\de\rho_Q\, ,\\
0 &=& \al_{p n}\de\rho_n+\al_{p p}\de\rho_p-2\al_{u u}\de\rho_u-\al_{d d}\de\rho_d\, ,\\
0 &=& \al_{n n}\de\rho_n+\al_{n p}\de\rho_p-\al_{u u}\de\rho_u-2\al_{d d}\de\rho_d\, .
\eq

Eqs.~(20)--(22) impose baryon number conservation, electric charge
neutrality and mechanical equilibrium as in the first case.
Eqs.~(23),(24) describe the chemical equilibrium respect to the two fast
processes of ``melting''.  As before we can calculate the chemical
imbalance associated to the weak reaction (19) and the corresponding
relaxation time. The rate of the process (19) is taken from Eqs.~(5)--(7) of
Ref.\cite{Sawyer:1989uy}(see also footnote
\footnote{The reaction rate has been multiplied by a factor three,
in agreement with Ref.\cite{Madsen:1992sx}.}).  The resulting bulk
viscosity is shown in Fig.~\ref{vis1} (lower panel).  The bulk viscosity of
``nucleon-quark'' MP is of the same order of magnitude of bulk
viscosity of strange matter \cite{Madsen:1992sx}.

\subsection{Effects of the surface tension}
Let us shortly discuss the effect of a non-vanishing surface tension
$\sigma$ at the interface between hadronic and quark matter. We will
assume that $\sigma\lesssim $ 30 MeV/fm$^2$, so that MP can form 
because it is energetically favored. 
Finite-size structures (drops, rods, slabs) form at different densities, 
to minimize the energy of the MP. 
In this case the perturbation
of the pressure due to r-modes is too weak to induce the ``melting''
process which is now a very slow process (in comparison to the
period of the
perturbation) and it plays no role in the calculation of the bulk
viscosity. 
The response of the finite size structures inside MP to a perturbation of the density 
corresponds to three possible processes:
a) the formation of a drop of ``new'' phase; b) the merging of two structures into
a single larger structure; c) the absorption of ``old'' phase into a structure of  
``new'' phase. Obviously, also the reverse processes are possible. It is easy to see that
large values of $\sigma$ suppress all these processes. In particular, 
in case a) the radius of a critical drop of new phase increases with $\sigma$, making it
more difficult to produce a new drop; case b) shares some similarity with the fission
problem in nuclear physics, where the process of the separation of a heavy nucleus into two
lighter nuclei is suppressed for larger values of $\sigma$ since, during the fission (or merging)
process, configurations having a large surface are produced. Finally, c) can be viewed
as a special case of b), in which the absorbed hadron can be assimilated to a small drop of quark matter.
Concerning the Coulomb interaction, it mainly plays a role in determining the size of the structures while
it is not so important for
the response of the structures to the perturbation, at least for relatively large values 
of the surface tension $\sigma \gtrsim $ 10 MeV/fm$^2$.
In that case, in fact, the screening due to electrons almost completely
cancels the effect of the Coulomb interaction in the nucleation rate \cite{Iida:1998pi}.
In the following for simplicity we will assume that 10 MeV/fm$^2$ $\lesssim \sigma \lesssim $
30 MeV/fm$^2$. 
In conclusion we reasonably assume that    
large values of $\sigma$ suppress all these processes,
therefore the equations describing the melting process as a 
fast reaction, i.e.
Eq.~(10) (first scenario) and Eqs.~(23), (24) (second
scenario) are not imposed.
We can compute the
viscosity by requiring that the baryon number and the electric charge
of the two phases are separately conserved.  This implies that Eq.~(6)
in the first scenario and Eqs.~(20), (21) in the second scenario, each
of them separate into two distinct equations. In the first
scenario, Eq.~(7) is not modified due to the charge neutrality of the
CFL quark phase. Finally, the equation of mechanical equilibrium 
(Eq.~(8) in the first scenario and Eq.~(22) in the second scenario)
is still valid and
it represents the only fast process connecting the two phases.  
Notice that since melting processes are suppressed, the only reaction
which allows the system to rapidly re-equilibrate is elastic scattering
between the two phases. In the following we have assumed that 
a residual interaction between the two phases always exists, 
similarly to the existence of 
``entrainment'' in superfluid neutron matter as discussed e.g. in Ref.\cite{Andersson:2002jd}.
The systems of equations read therefore:
\bq
0 &=& (1-\chi)(\de\rho_n+\de\rho_p+\de\rho_\Lambda+\de\rho_\Sigma)\nonumber\\
  &-&\de\chi(\rho_n+\rho_p+\rho_\Lambda+\rho_\Sigma)\, ,\\
0 &=& \chi\de\rho_q+\de\chi\rho_q\, ,\\
0 &=& (1-\chi)(\de\rho_p-\de\rho_\Sigma)-\de\chi(\rho_p-\rho_\Sigma)\, ,\\
0 &=& \sum_{\{H\}}p_H\de\rho_H- p_q\de\rho_q\, ,\\
0 &=& \beta_n\de\rho_n+\beta_p\de\rho_p+\beta_\Lambda\de\rho_\Lambda+\beta_\Sigma\de\rho_\Sigma\, \label{hyp}
\eq
for the first scenario and

\bq
0 &=& (1-\chi)(\de\rho_n+\de\rho_p)-\de\chi(\rho_n+\rho_p)\, ,\\
0 &=& \chi(\de\rho_u+\de\rho_d+\de\rho_s)+\de\chi(\rho_u+\rho_d+\rho_s)\, ,\\
0 &=& (1-\chi)\de\rho_p -\de\chi\rho_p\, ,\\
0 &=& \chi(2\de\rho_u-\de\rho_d-\de\rho_s)+\de\chi(2\rho_u-\rho_d-\rho_s)\\
0 &=& \sum_{\{H\}}p_H\de\rho_H-\sum_{\{Q\}}p_Q\de\rho_Q\, 
\eq
for the second scenario.

%%%%%%%%%%%%%%%%%%%%%%%%%%%%%%%%%%%%%%%%%%%%%%%%%%%%%%%%%%%%%%%%%%%%%

\begin{figure}[]
\begin{center}
\includegraphics[scale=0.55]{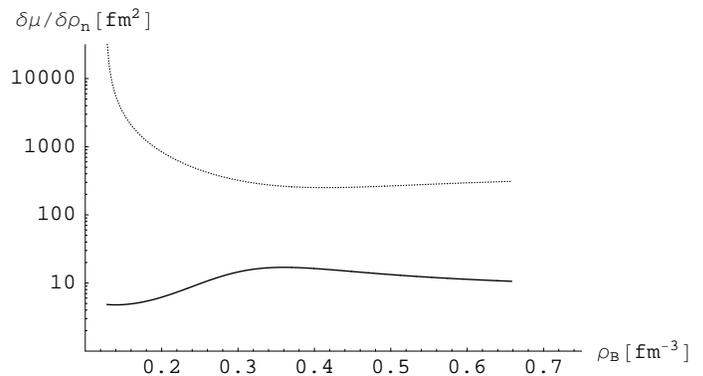}
\end{center}
\parbox{7.5cm}{
\caption{\label{unbalance}The chemical unbalance for the second scenario 
is shown as a function of the baryonic density.
The solid line corresponds to the case of a vanishing of the surface tension and 
the dotted line corresponds to a finite surface tension.
}}
\end{figure}

%%%%%%%%%%%%%%%%%%%%%%%%%%%%%%%%%%%%%%%%%%%%%%%%%%%%%%%%%%%%%%%%%%%%
In Fig.~\ref{vis1} and \ref{vis2} we show the effect of a non-vanishing surface
tension on the viscosity (dotted lines). On rather general grounds,
one can expect that the effect of a non-vanishing surface tension is
to reduce the viscosity. Indeed, the surface tension suppresses
the fast processes of ``melting'' which are responsible for
the reduction of the chemical unbalance. In particular, the only
equation connecting the two phases is now the equation corresponding
to the mechanical equilibrium. Moreover, in the first scenario the
number of constraints on $\de \rho_H$ reduces from four to three and,
in the second scenario, the constraints on
$\de\rho_Q$ from four to two. It is also interesting to 
notice that in both scenarios, near the beginning of the
MP ($\chi \rightarrow 0$),  baryon number conservation requires $\de \chi
\rightarrow 0 $ if the surface tension is non-vanishing (see Eqs.~(26) and (31)).  
On the other hand, in the first scenario $\rho_\Lambda=0$  
at the beginning of the MP below the
$\Lambda$ production threshold,
and therefore, the constraint $\de \chi =0$ is also satisfied 
in the absence of surface tension, as explained in footnote [32]. Therefore the dotted lines
coincide with the solid lines in both panels of Fig.~\ref{vis1} for
$\rho_B \lesssim 0.6$ fm$^{-3}$. 

In the second scenario, the effect
of the surface tension is more dramatic. In Fig.~\ref{unbalance} we display 
the chemical unbalances corresponding to a vanishing and to a finite value of the surface
tension, respectively. As already remarked $\de \mu/\de \rho_n$ is larger in presence of
a surface tension. Concerning the singular behavior of the chemical
unbalance near the first critical density, it stems from neglecting
$\de \rho_e$ in the equation of the electric charge
conservation. While in general this is a safe approximation 
 due to the slowness of the modified Urca
process (see Ref.~\cite{Lindblom:2001hd}), in this particular case this approximation implies the
vanishing of the viscosity at threshold. Actually, the existence of a
finite value for $\de\rho_e$ implies that the viscosity is small but
finite at threshold. It is not possible to apply directly the formalism 
of Ref.~\cite{Lindblom:2001hd} if two {\it independent } perturbations
($\de\rho_n$ and $\de\rho_e$) exist in the system. We are therefore forced
to discuss separately the viscosities stemming from the two independent
unbalances. Concerning the viscosity associated with the modified Urca process,
we remind that it scales as $T^6$ and therefore it is essentially negligible
below $T \sim 10^{10}$ K while it will be included in the calculation of the 
stability of the star presented in the next section.
In conclusion, the result corresponding to the dotted line of Fig.~\ref{vis2}
would not be significantly modified by taking into account a finite value of  $\de\rho_e$. 

%%%%%%%%%%%%%%%%%%%%%%%%%%%%%%%%%%%%%%%%%%%%%%%%%%%%%%%%%%%%%%%%%%%%%

\begin{figure}[t]
\begin{center}
\includegraphics[scale=0.55]{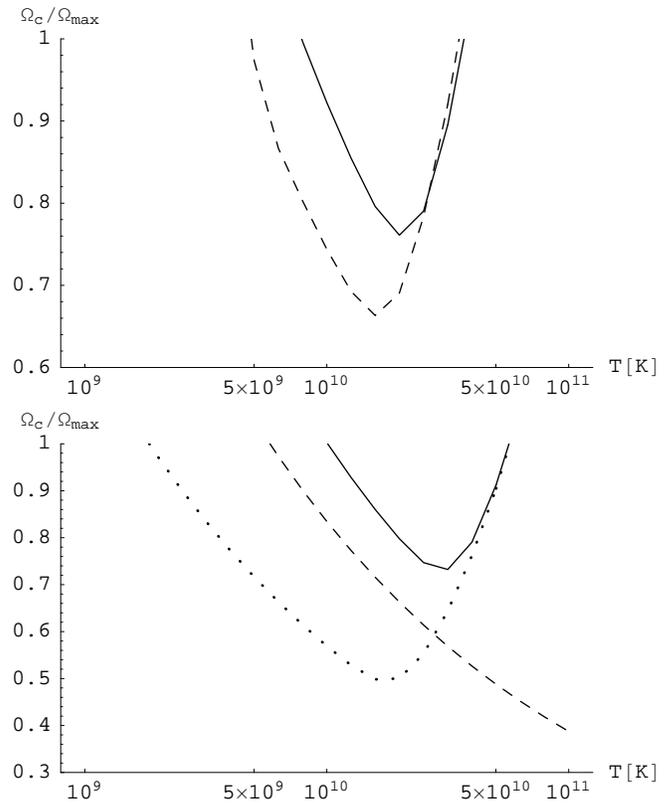}
\end{center}
\parbox{7.5cm}{
\caption{\label{stabilita}
Critical angular velocities. The solid lines refer to
hybrid stars, dashed lines correspond to hyperonic stars (upper panel)
and to strange stars (lower panel). The dotted line 
corresponds to the case of a large hadron-quark 
surface tension (10 MeV/fm$^2$  $< \sigma <$ 30 MeV/fm$^2$, see text
for details).}  }
\end{figure}

%%%%%%%%%%%%%%%%%%%%%%%%%%%%%%%%%%%%%%%%%%%%%%%%%%%%%%%%%%%%%%%%%%%%

\section{Stability of Hybrid Stars}
We can now address the problem of the stability of a rotating compact
star.  To compute the critical angular velocity we use the standard
formalism of
Refs.\cite{Lindblom:1998wf,Lindblom:1999yk,Lindblom:2001hd}.  We need
first to integrate the viscosity on the structure of the star, which
is obtained by solving Tolman-Oppenheimer-Volkov equation.  In Fig.~\ref{profili}
we show the structure of a $M=1.46 M_\odot$ star for the two
scenarios discussed above. 
For simplicity we computed the structure of the star assuming a negligible
value for $\sigma$. The main effect of the presence of finite size 
structures is to reduce the volume occupied by the MP. For $\sigma \lesssim$
30 MeV/fm$^2$, which is the limit of validity of our approach,
the shrinking of the MP is rather modest \cite{Heiselberg:1992dx}.   
The critical angular velocity
$\Omega_{crit}$ is the one for which the imaginary part of the r-mode
frequency vanishes and it is obtained by solving the equation
\be
-1/\tau_{GR}+1/\tau_{B}+1/\tau_{B(Urca)}=0 \, .
\ee
Here $\tau_{GR}$ is the time scale for gravitational waves emission while
$\tau_{B}$ and $\tau_{B(Urca)}$ are the time scales of the bulk viscosity  
produced by hadronic processes and by the modified Urca
process of the nucleons\cite{Sawyer:1989}, respectively.  Results 
for the critical angular velocity are shown in
Fig.~\ref{stabilita}.  In the upper panel we compare the stability of a purely
hyperonic star with the stability of a HyS containing
hyperon-quark MP. As it can be seen, due to the large viscosity of the
MP the HyS is as stable as the hyperonic star. Let us stress
again that our result indicate that the viscosity in the MP is almost
independent on the hyperonic content and therefore hybrid
hyperon-quark stars can be stable as long as a tiny fraction of
hyperon is present.  In the lower panel we compare the stability of a
star made entirely of non-interacting quarks with a HyS containing
a nucleon-quark MP. We have assumed that the
viscosity of the pure quark matter phase in the HyS vanishes, to simulate a CFL
core. As it can be seen, the small MP region, located near the edge of
the star, is sufficient to damp the r-modes. For simplicity, we have
assumed quarks to be unpaired in the MP, but similar results should be
obtainable if a 2SC phase is present. 
A feature of r-modes is that they are active mostly in the
outer regions of the star, and therefore the value of the bulk
viscosity at a not too large densities is crucial for the stability of the
star. In particular, in the first scenario 
the stability of the star is due to the
presence of the
$\Sigma$ hyperons in the pure hadronic phase and in the MP.  $\Lambda$
hyperons, which are produced at larger densities, play a lesser role.  
In the second scenario, a small window of MP, present in the outer region, is sufficient to stabilize the star.
Let us also remark that the
star is stable at large temperatures, due to the modified Urca
processes active in the crust. These processes does not exist in the case of pure quark stars,
which are therefore unstable at large temperatures.
Finally, when a finite value for the hadron-quark surface tension is taken into account
the instability window is larger,
but the main conclusion concerning the stability of a young hybrid star
remains valid.

It is a pleasure to thank J.C.~Miller and L.~Rezzolla for very useful
discussions.

\bibliography{references}
\bibliographystyle{apsrev}
\end{document}